\documentclass{article}

\usepackage[utf8]{inputenc}

\usepackage[normalem]{ulem}
\usepackage{caption}
\usepackage{xcolor}
\usepackage{graphicx}
\usepackage{float}
\usepackage{authblk}
\usepackage{amsmath}
\usepackage{amssymb}
\usepackage{setspace}
\usepackage{mathtools}
\usepackage{booktabs}
\usepackage{hyperref}

\usepackage{multirow}%
\usepackage{amsfonts}%
\usepackage{amsthm}%
\usepackage{mathrsfs}%
\usepackage[title]{appendix}%
\usepackage{textcomp}%
\usepackage{manyfoot}%
\usepackage{algorithm}%
\usepackage{algorithmicx}%
\usepackage{algpseudocode}%
\usepackage{listings}%

\title{Ultrafast room temperature valley manipulation in silicon and diamond}

\author[1]{Adam Gindl}
\author[1]{Martin Čmel}
\author[1]{František Trojánek}
\author[1]{Petr Malý}
\author[1,*]{Martin Kozák}

\date{}

\affil[1]{Department of Chemical Physics and Optics, Faculty of Mathematics and Physics, Charles University, Ke Karlovu 3, 12116 Prague 2, Czech Republic}

\affil[*]{Corresponding author: m.kozak@matfyz.cuni.cz}

\begin{document}

	\maketitle
	
	\section*{Abstract}
	\textbf{Some semiconductors feature more than one degenerate minimum of the conduction band in their band structure. These minima – known as valleys – can be used for information storage and processing if it is possible to generate a difference in their electron population. However, to compete with conventional electronics, it is necessary to develop universal and fast methods for controlling and reading the valley quantum number of electrons. Even though selective optical manipulation of electron populations in inequivalent valleys has been demonstrated in two-dimensional crystals with broken time-reversal symmetry, such control is highly desired in many technologically important semiconductor materials including silicon and diamond. Here, we demonstrate an ultrafast technique for the generation and read-out of a valley-polarized population of electrons in bulk semiconductors on sub-picosecond time scales. The principle is based on unidirectional intervalley scattering of electrons accelerated by oscillating electric field of linearly polarized infrared femtosecond pulses. Our results are an advance in the development of potential room temperature valleytronic devices operating at terahertz frequencies and compatible with contemporary silicon-based technology.   }

	\section*{Main text}
	
	In crystalline materials, the electron wave function follows the periodicity of the lattice and can be written as a Bloch wave $\psi_{n \mathbf{k}}(\mathbf{r})=u_{n \mathbf{k}}(\mathbf{r})e^{i\mathbf{k.r}}$, where $u_{n \mathbf{k}}(\mathbf{r})$ is the spatially periodic part and $\mathbf{k}$ is the wave vector of the electron. The solution of time-independent Schrödinger equation leads to electronic states in the form of continuous bands with energy dispersion $E_n(\mathbf{k})$, where $n$ is the band index. Semiconductors and dielectric materials are characterized by a presence of a band of forbidden energies separating the highest occupied valence band and the lowest unoccupied conduction band. 
	
	The dispersion relation $E_c(\mathbf{k})$ of the conduction band possesses either a single minimum in the center of the first Brillouin zone ($\mathbf{k}=0$) or several energy degenerate minima (valleys) corresponding to different values of the wave vector $\mathbf{k}$. In materials with multiple conduction band valleys, the quantum number associated with the wave vector $\mathbf{k}$ of the occupied valley can be utilized for transport, processing and storage of information instead of the electric charge of electron, which is exploited in classical electronics.
	
	In specific two-dimensional crystals with broken time-reversal symmetry (e.g. 2D transition metal dichalcogenides \cite{Mak2010,Splendiani2010}), spin-orbit coupling leads to existence of two inequivalent groups of energy degenerate valleys with opposite electron spin \cite{Xiao2012}. The excitons can be populated to a specific group of valleys thanks to the selection rules for optical transitions induced by circularly polarized light \cite{Yao2008,Cao2012,Wang2018}. The valley quantum number can be selected by the handedness of the circular polarization of resonant photons generating the excitons in these materials. However, in bulk semiconductors with multiple valleys such as silicon or diamond, such selection rules do not exist due to the crystal symmetry. Alternative ways of manipulation with specific conduction band valleys are based on removing the energy degeneracy by static electric or magnetic field \cite{Aivazian2015,Stier2018,Molas2019, Cunningham2019, Slobodeniuk2023}, by spatial confinement of electrons \cite{Takashina2006}, via optical coherent phenomena \cite{Kim2014,Sie2015,Sie2017} or by nonresonant strong-field interaction with shaped intense fields \cite{Langer2018,Tyulnev2024}.
	
	A conceptually different approach has been developed to generate, transport and detect valley polarized population of electrons in diamond at low temperatures \cite{Isberg2013,Suntornwipat2021}. The approach is based on application of a strong static electric field along [100] crystallographic direction. Due to anisotropic acceleration of electrons in inequivalent valleys caused by anisotropy of the effective mass tensor, the intervalley electron-phonon scattering rate is higher for the process, in which the electrons scatter from a valley with small effective mass in the direction of the applied field to a valley with large effective mass. The populations of electrons from different valleys are detected electrically by time-of-flight measurement of electrical current and can be separated by application of magnetic field in Hall effect measurements \cite{Isberg2013}. This method of achieving and detecting valley polarization requires the anisotropic electron distribution to persist in the crystal for several tens of nanoseconds due to the necessity of charge transport over macroscopic distance and its collection by the electrodes. The valley-polarized electron population relaxes to isotropic distribution in reciprocal space via intervalley electron-phonon scattering. The change in the wave vector of the electron $\Delta \mathbf{k}$ during its transition to another valley is compensated by the creation/annihilation of a phonon, which has a wave vector $\mathbf{q}=-\Delta \mathbf{k}$. While the characteristic intervalley scattering time of electrons in semiconductors and dielectric crystals at cryogenic temperatures is up to 1 millisecond, it decreases to femtosecond to picosecond time scales at room temperature \cite{Hammersberg2014,Ichibayashi2011}. For this reason, the valley polarization of electrons has not been observed in bulk semiconductors or dielectric crystals at room temperature because it requires an ultrafast technique both for its generation and detection.

	Here we show that valley polarization of electrons in silicon and diamond crystals can be induced by using oscillating electric field of femtosecond infrared pulses. These two covalent semiconductors share the same double face-centered cube structure leading to similar properties of the conduction band dispersion, although the band gap energies significantly differ (1.12 eV in silicon, 5.5 eV in diamond). Both materials have indirect band structure and their conduction band dispersion $E_c(\mathbf{k})$ possesses six energy degenerate minima located close to the X-point at the Brillouin zone edge in [100] and equivalent directions. The first Brillouin zone of silicon/diamond crystals is shown in Fig. 1a along with the surfaces of constant energy for the six degenerate valleys. In parabolic approximation, the energy of an electron in $i$-th valley can be written as $E_c^{(i)}(\mathbf{k})=E_\text{g}+\hbar^2/2 \sum_{j} (k_j-k_{j,0}^{(i)})^2/m^{(i)}_j $, where $E_\text{g}$ is the band gap energy, $k_j$ denotes the individual components of the electron wave vector, $k_{j,0}^{(i)}$ are the coordinates of the local band minimum and $m^{(i)}_j=\hbar^2\left[ \partial^2 E_c^{(i)}(k_j)/\partial k_j^2 \right]^{-1}$ is the electron effective mass. The electron effective mass tensor in each valley is strongly anisotropic with the longitudinal effective mass $m_\text{l}$ being approximately 5-times larger than the transverse effective mass $m_\text{t}$ both in silicon and diamond \cite{Dexter1954,Naka2013}. There are three inequivalent groups of valleys with large effective mass in crystallographic directions [100] (along $\mathbf{k}_x$), [010] (along $\mathbf{k}_y$) and [001] (along $\mathbf{k}_z$). When we apply an oscillating electric field in [100] direction, an electron with the initial momentum $\mathbf{k}(0)=\mathbf{k_{j,0}}$ (center of one of the valleys) is accelerated in the $\mathbf{k}_x$ direction in the momentum space and its oscillations follow the pump field. While the excursion of the electron in momentum space $\mathbf{k}_\text{max}=(e\mathbf{F}_0)/(\hbar\omega)$ does not depend on the effective mass, the average kinetic energy of an oscillating electron $U_{\text{p}}=  (e|\mathbf{F}_0|)^2/(4\omega^2m_j^{(i)})$ differs depending on which valley the electron occupies due to the effective mass anisotropy. Here $\mathbf{F}_0$ is the field amplitude of the pump pulse, $e$ is electron charge and $\omega$ is the angular frequency of the pump field.
	
	The electrons exchange energy and momentum with the crystal via the interaction with phonons. The mechanism of electron transfer between inequivalent groups of valleys is based on intervalley electron-phonon scattering (in the literature referred to as f-scattering). To conserve momentum, such transition requires absorption or emission of a phonon with large momentum. Specifically, the f-scattering between inequivalent valleys in silicon and diamond proceeds via the interaction of electrons with longitudinal and transverse acoustic and transverse optical phonons \cite{Jacoboni1983}. Equilibrium population of these modes at room temperature is low due to relatively high energies of phonons with large momenta (see Table 1 in Methods). As a result, most of the intervalley transitions occur via phonon emission, which is only possible for electrons with kinetic energy larger than the energy of the phonon involved in the scattering process. The dependence of the intervalley electron-phonon scattering rate on the electron energy combined with different values of the average kinetic energy of electrons in different groups of valleys during the interaction with the electric field of the infrared pump pulse leads to a higher probability of intervalley transitions of electrons from the valleys with low effective mass $m_\text{t}$ in the direction of the applied oscillating electric field (red ellipsoids in Fig. 1b) to the valleys with large effective mass $m_\text{l}$ (blue ellipsoids in Fig. 1b) than in the opposite direction. Although the unidirectional intervalley electron transitions are present only for a short duration of the infrared pulse, a significant fraction of the electrons is transferred due to the high intervalley scattering rate of high energy electrons of up to $\approx 10^{13}$ s$^{-1}$.
	
	To verify that this mechanism can transfer electrons between inequivalent conduction band valleys within few tens of femtoseconds we solve Boltzmann transport equation using Monte-Carlo approach closely following the treatment in \cite{Jacoboni1983}. In the simulations we consider the intervalley scattering due to the interaction of electrons with acoustic and optical phonons and intravalley acoustic phonon scattering. The electron-phonon scattering is described in the framework of deformation potential with the scattering rate dependent on the electron momentum and lattice temperature (details of simulations are described in Methods section). The calculated time evolution of electron populations in the three inequivalent groups of valleys with the longitudinal effective mass along [100], [010] and [001] directions during and after the interaction with the infrared femtosecond pump pulse is shown in Fig. 1c for silicon and in Fig. 1d for diamond. Both simulations assume room temperature of the crystal and pump pulse properties, which are applied in the following experiments. In silicon, up to 40\% of electrons are transferred to the two valleys with the longitudinal effective mass along the pump polarization after the interaction with the infrared pump pulse with photon energy of 0.62 eV, the duration of 40 fs and the peak electric field $F_0$=0.7 V/nm. We can define the degree of valley polarization as $V=\left ( N_{[100]}-N_{[010]} \right )/N$, where $N_{[100]}$ and $N_{[010]}$ are the volume densities of electrons in valleys with the direction of large effective mass along [100] and [010] crystallographic directions, respectively, and $N$ is the total electron density in the conduction band. Under the conditions described above, the degree of valley polarization in silicon obtained from the numerical simulations is $V \approx 0.10$. In diamond, the population of the two valleys with large effective mass along pump polarization increases up to 55\% of the total electron population for the field amplitude of the pump pulse of $F_0$=1.3 V/nm corresponding to $V \approx 0.33$. These values are comparable to the degree of valley polarization of 32\% obtained from helicity dependent photoluminescence measurements in 2D transition metal dichalcogenide MoS$_2$ excited by circularly polarized light  \cite{Zeng2012}. Calculated electron distribution in momentum space during and shortly after the interaction with the laser pulse is shown in Supplementary Video 1 and 2 for both materials.
	
	The valley polarization induced in silicon and diamond is detected via polarization anisotropy of free carrier absorption of a probe pulse. In the infrared spectral region, the difference between excited carrier absorption coefficients $\alpha_{[100]}$ and $\alpha_{[010]}$ for probe polarization components along [100] and [010] directions, respectively, can be approximated using Drude model as \cite{Yu1999}:
	
	\begin{equation}
		\label{eq_abs}
		{\Delta\alpha=\alpha_{[100]}-\alpha_{[010]}=A\frac{(m_\text{t}-m_\text{l})(N_{[100]}-N_{[010]})}{m_\text{l} m_\text{t}},}
	\end{equation} where $A$ is a constant (see Methods for details). We note that holes do not contribute to the measured $\Delta \alpha$ because their effective mass tensor is symmetric with respect to rotation by 90$^\circ$. The degree of valley polarization is directly proportional to the measured $\Delta \alpha$ via Eq. (\ref{eq_abs}) and it can be expressed as $V=\Delta \alpha m_l m_t/\left [ AN(m_t-m_l) \right ]$. 
	
	Ultrafast optical generation and detection of valley polarized electron population in both studied materials is experimentally demonstrated using the scheme shown in Fig. 1a, b (detailed layout of the experimental setup is shown in Supplementary Figure 1). First the electrons and holes are excited by a pre-excitation pulse (Fig. 1a). To excite the electrons with homogeneous distribution along the light propagation we use indirect single-photon excitation in silicon (photon energy 1.2 eV) and two-photon excitation in diamond (photon energy 3.6 eV). Several picoseconds after the optical excitation, the electrons relax to the band minima via electron-phonon scattering and distribute isotropically to the six degenerate conduction band valleys (blue elipsoids in Fig. 1a). In the time delay of 100 ps after photoexcitation, the crystal is illuminated with the infrared pump pulse with linear polarization along [100] or [010] direction, which induces the valley polarization. The waiting time of 100 ps is selected to ensure that the electronic system is already relaxed to the lattice temperature when the pump pulse arrives to the sample. We note that this time delay is significantly shorter than the lifetime of excited carriers in both materials \cite{Schroder1997,Kozak2014}. The anisotropy of [100] and [010] polarization components of free carrier absorption is measured using a probe pulse with linear polarization rotated by 45$^\circ$ with respect to the polarization of the pump (along [-110] direction), which is incident on the sample in the time delay $\Delta t$ after the pump pulse. Both the pump and probe pulses have photon energy of 0.62 eV (central wavelength of 2000 nm) and pulse duration of 40 fs. The peak electric field of the pump pulse is up to 1.3 V/nm in the experiments performed in silicon and 1.5 V/nm in diamond (details of the experimental setup are discussed in Methods).
	
	The measured polarization anisotropy of free carrier absorption $\Delta \alpha$ at room temperature is shown in Fig. 1e, f for both studied materials with the polarization of the pump along [100] (black curves) and [010] (red curves) directions. Different sign of $\Delta \alpha$ for the two orthogonal directions of pump polarization confirms that the electrons are transferred preferentially to the valleys with the longitudinal effective mass along the direction of pump polarization. In the following experiments, the difference between the measured $\Delta \alpha$ with pump polarization along [100] and [010] is used to remove any possible experimental artifacts which could influence the measurement of the polarization anisotropy of free carrier absorption. The fast signal at $\Delta t=0$ ps is caused by the coherent interaction of pump and probe pulses. The valley polarized electron population is proportional to the slow exponentially decaying component of $\Delta \alpha$.
	
	To verify the origin of nonzero $\Delta \alpha$ in valley polarized electron population we rotate both crystals by 45$^\circ$ such that the pump polarization is along [110] or [1-10] (see Extended Data Figure 1). In this configuration we do not observe any decaying $\Delta \alpha_{[110]}=\alpha_{[110]}-\alpha_{[1-10]}$ because the four valleys with the longitudinal effective mass perpendicular to [001] (blue ellipsoids in the upper insets of Extended Data Figure 1a,b) are populated isotropically for pump polarization along [110] and [1-10].

	To generate valley polarized electron population, the amplitude of the electric field of the pump pulse has to be adjusted such that the maximum of the instantaneous energy of electrons in the valleys with low effective mass becomes higher than the energy of phonon required for intervalley scattering while the energy of the electrons from the valleys with large effective mass has to stay below the phonon energy. The measured dependence of the slow component of $\Delta \alpha$ and the corresponding degree of valley polarization of electrons on the electric field amplitude of the pump pulse is shown in Extended Data Figure 2 for both materials compared to numerical simulations. The degree of the induced valley polarization saturates and even decreases for excessively large pump field. The difference between the measured data and the numerical calculations in silicon at field amplitudes $F_0>0.7$ V/nm is caused by the fact that the photon energy of the pump is close to the silicon band gap energy. The nonresonant pump induces transitions to higher conduction bands, from which the electrons relax and distribute isotropically to the six valleys. This effect decreases the degree of valley polarization at pump field $F_0>0.7$ V/nm. The discrepancy between the measured and calculated degree of valley polarization in diamond at field amplitude $F_0>1.3$ V/nm is probably due to the band nonparabolicity, which is not taken into account in the numerical simulations. The generated degree of valley polarization can be further optimized by modifying the photon energy, duration and peak field of the pump pulses. Numerical simulations at crystal temperature of 77 K shown in Extended Data Figure 3 suggest that when decreasing the frequency of the pump pulse to THz region, the maximal degree of valley polarization increases up to $V=$0.35 in silicon and $V=$0.92 in diamond. We note that the amplitude of the oscillating field required to generate valley polarized electron population is about 4 orders of magnitude higher than the amplitude of the DC field required to reach valley polarization in diamond at low temperatures \cite{Isberg2013} and it strongly depends on the frequency of the pump pulse (see Extended Data Figure 3).
	
	Compared to previous experiments investigating the valley polarization of electrons in silicon \cite{Takashina2006,Sakamoto2013} or diamond \cite{Isberg2013}, our method offers femtosecond time resolution allowing us to directly characterize the relaxation dynamics of valley polarization due to intervalley electron-phonon scattering in both studied crystals. The measured relaxation times of valley polarization are shown in Fig. 2a,b as a function of sample temperature for electron density $N=1.8\times10^{17}$ cm$^{-3}$ (black squares in Fig. 2a) and $N=1.5\times10^{18}$ cm$^{-3}$ (green circles in Fig. 2a) in silicon and $N=4.4\times10^{15}$ cm$^{-3}$ (black squares in Fig. 2b) and $N=6.4\times10^{16}$ cm$^{-3}$ (green circles in Fig. 2b) in diamond. Carrier density is controlled by the fluence of the pre-excitation pulse. At the lowest carrier densities used in our experiments, the measured relaxation times of valley polarization at room temperature are $\tau_\text{rel}=730$ fs in silicon and $\tau_\text{rel}=9.7$ ps in diamond. The former value can be compared to the L-X intervalley scattering time of 180 fs reported for electron in silicon in \cite{Ichibayashi2011}. In both materials we observe that the relaxation time of valley polarization increases with decreasing temperature in agreement with numerical calculations of intervalley scattering time (solid curves in Fig. 2a,b, details in Methods) reaching $\tau_\text{rel} \approx 80$ ns in silicon and $\tau_\text{rel} \approx 10$ ns in diamond. Valley polarized population of electrons thus persists in these materials at low temperatures for significantly longer times than in 2D transition metal dichalcogenides, in which the intervalley scattering times of only few picoseconds were reported \cite{DalConte2015}. While the measured relaxation time depends strongly on the excited carrier density in the diamond crystal at low temperatures, the dynamics in silicon practically does not change when changing the pre-excited electron density. This is due to Coulomb interaction and many-body phenomena leading to formation of exciton molecules and electron hole droplets with different critical temperature in diamond $T_\text{c}=$165 K \cite{Shimano2002,Kozak2013} and silicon $T_\text{c}=$27 K \cite{Combescot1974,Shah1977}. Below these temperatures the electron-hole condensation leads to Auger recombination. In this nonradiative process, an electron-hole pair recombines while transferring the energy to another quasiparticle (electron or hole). Compared to previous experiments in diamond, in which the valley relaxation time of 300 ns was reported at 77 K with very low excited carrier density of below $N=10^{10}$ cm$^{-3}$ \cite{Isberg2013}, the relaxation of valley polarization at low temperatures is in our case (excited carrier density $N=10^{15}-10^{17}$ cm$^{-3}$) significantly influenced by faster carrier recombination and by higher intervalley scattering rate of electrons, which increases due to the excess energy obtained during Auger process.

	Even at moderate excited electron densities of the order of $N=10^{13}-10^{15}$ cm$^{-3}$, the measured relaxation time of valley polarization in diamond is influenced by Coulomb interaction. Due to the high binding energy of Wannier excitons of 80 meV \cite{Dean1965}, most of the excited carriers are bound in the excitonic state at low temperatures (the balance between excitons and free electrons and holes can be described using Saha equation \cite{Gourley1982,Ichii2020}). The presence of excitons influences the processes of generation and relaxation of valley polarization via an increased deformation potential of exciton state compared to the single electron state.

	The possibility to switch the valley polarization on femtosecond time scales can be utilized in future valleytronic devices working at THz frequencies. To demonstrate the ultrafast capabilities of our method for generation, switching and readout of valley polarization of electrons we illuminate the silicon and diamond crystals by a pair of infrared pump pulses separated in time. The first pump pulse is linearly polarized parallel to [100] direction and generates the valley polarization with most of the electrons located in the valleys with their longitudinal effective mass along [100] leading to a decrease of the measured $\Delta \alpha$. The second pump pulse arriving to the sample in the time delay of 1.4 ps with respect to the first pump is linearly polarized along [010] direction (see Methods for details of pulse pair generation), rotates the direction of valley polarization and thus the sign of the measured $\Delta \alpha$ (see Fig. 3a for the measurement in silicon at temperature of 7 K and Fig. 3b for the measurement in diamond at room temperature). The time needed for valley polarization switching and its observation is ultimately limited by the duration of the pump and probe pulses and by electron relaxation in the valleys. In silicon, the time resolution is also limited by the coherent interaction of the pump and probe pulses inducing two-photon transitions to higher conduction bands (decrease of $\Delta \alpha$ in time delay $\Delta t=0$ ps and its increase in $\Delta t=1.4$ ps). In both materials we clearly reach sub-picosecond switching of valley polarization of electrons.

	There is a large number of materials in which the valley polarization of electrons has never been observed due to the lack of an ultrafast method, which allows to induce and detect the anisotropic electron population in momentum space before it relaxes back to the isotropic state. Due to its nonresonant nature, the technique presented here can be applied to generate valley polarized electron population in such materials with only two requirements: i) the effective mass tensor of the electrons has to have different anisotropy in inequivalent valleys and ii) the intervalley scattering rate has to depend on the electron energy in the band. However, these two conditions are generally valid for any semiconductor or dielectric material with multiple degenerate minima of the conduction band. The principles demonstrated here thus open the field of valleytronics to a wide range of crystals and possibly also to nanocrystalline materials.
	
	\section*{Acknowledgements}
	
	The authors would like to acknowledge the support by Charles University \\(UNCE/SCI/010, SVV2020-260590, PRIMUS/19/SCI/05, GA UK 124324), by the Czech Science Foundation (23-06369S) and by the European Union (ERC, eWaveShaper, No. 101039339). Views and opinions expressed are, however, those of the author(s) only and do not necessarily reflect those of the European Union or the European Research Council. Neither the European Union nor the granting authority can be held responsible for them. This work was supported by TERAFIT project No. CZ.02.01.01/00/22\_008/0004594 funded by OP JAK, call Excellent Research.
	
	\section*{Author contributions}
	
	M.K. conceived and designed the study and performed the numerical simulations. A.G., M.Č. and M.K. performed the experiments and analysed the data. M.K. wrote the manuscript with the contribution from A.G., M.Č., F.T. and P.M.

	\section*{Ethics declarations}
	
	\subsection*{\label{sec:level2}Competing interests}
	The method for ultrafast storage and readout of information into the valley polarized electron population in bulk crystals described in this manuscript is subject of patent application number PV 2024-273. Patent applicant is Charles University. Names of inventors are M. Kozák, M. Čmel, A. Gindl, F.Trojánek and P. Malý. The application was submitted on July 4th 2024.

		\section*{Figures:}
		
	\begin{figure}[H]
		\center
		\includegraphics[width = 1\linewidth]{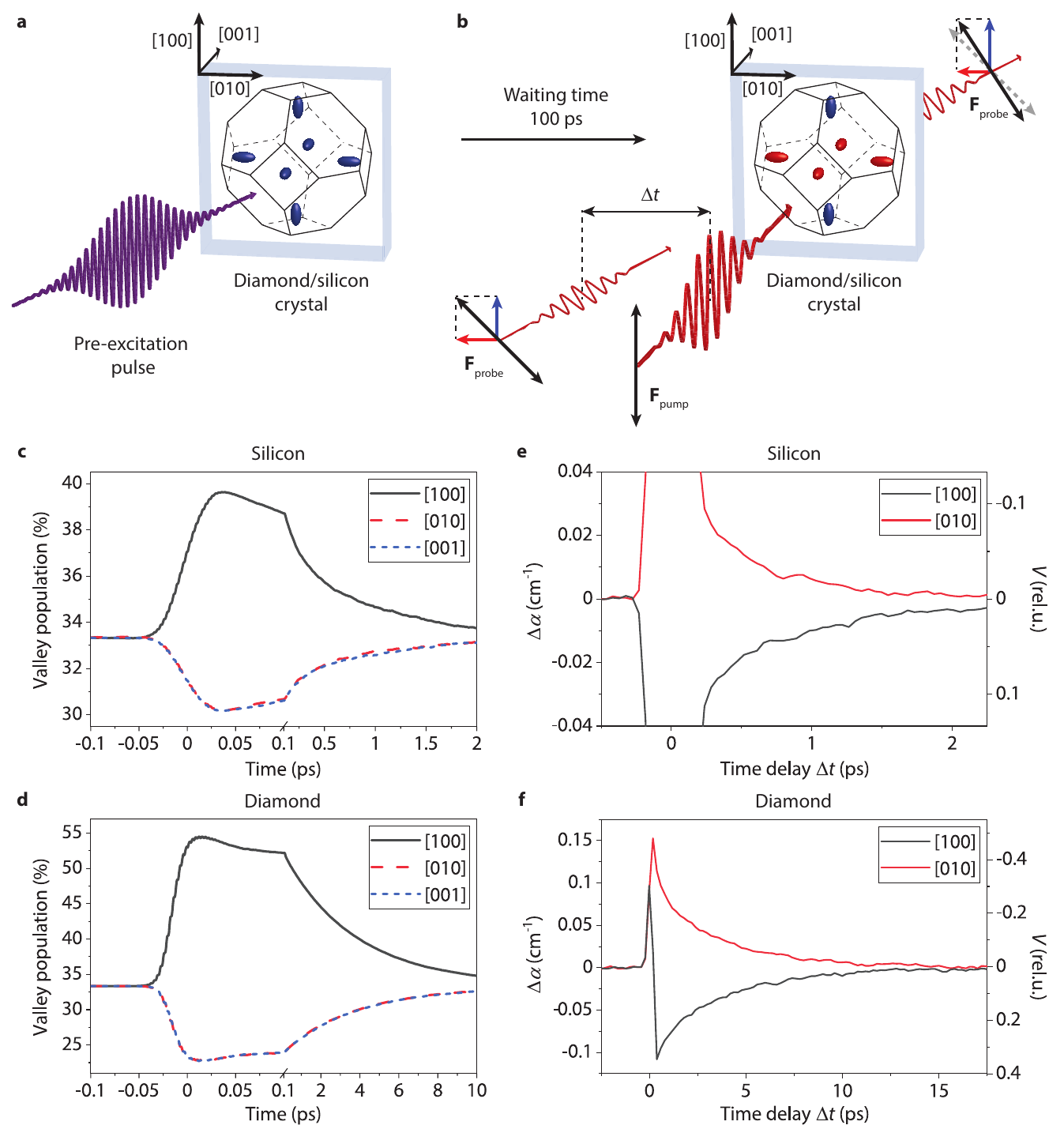}
		\captionsetup{labelformat=empty}
		\caption{\textbf{Fig. 1. Generation and detection of valley polarized electron population in silicon and diamond. a} The electrons are excited by a resonant pre-excitation pulse and equally distributed to all conduction band valleys (blue ellipsoids represent constant energy surfaces). \textbf{b} After 100 ps, the linearly polarized infrared pump pulse generates the valley polarized electron distribution with higher population in valleys with their principal axes parallel to the pump polarization (blue ellipsoids) than in the other valleys (red ellipsoids). Valley polarization is measured via polarization anisotropy of free carrier absorption of a linearly polarized probe pulse with polarization rotated by 45$^\circ$ incident on the sample in time delay $\Delta t$ with respect to the pump pulse. \textbf{c, d} Time evolution of electron populations in the three inequivalent groups of valleys in silicon (\textbf{c}) and diamond (\textbf{d}) at room temperature calculated by Monte-Carlo simulations. \textbf{e, f} Measured polarization anisotropy of free carrier absorption and the corresponding degree of valley polarization $V$ in silicon (\textbf{e}) with pre-excited electron density of $N=1.8 \times 10^{17}$ cm$^{-3}$ and diamond (\textbf{f}) with $N=6.4 \times 10^{16}$ cm$^{-3}$ at room temperature for two orientations of the pump polarization along [100] direction (black curve) and [010] direction (red curve). Electric field amplitude of the pump pulse is 0.7 V/nm in silicon and 1.3 V/nm in diamond both in experiment and simulation.}
		\label{Fig1}
	\end{figure}

	\begin{figure}[H]
		\center
		\includegraphics[width = 1\linewidth]{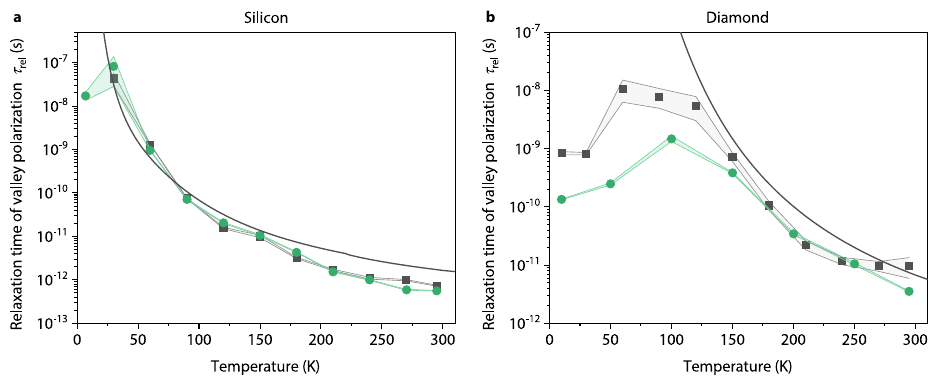}
		\captionsetup{labelformat=empty}
		\caption{	\textbf{Fig. 2. Relaxation time of valley polarization in silicon and diamond crystals as a function of lattice temperature.} Measured relaxation time of valley polarization $\tau_\text{rel}$ in silicon \textbf{a} and diamond \textbf{b} compared with the intervalley scattering time obtained from numerical solution of Boltzmann transport equation using Monte-Carlo approach (solid curve). The experimental points represent the decay time obtained by fitting the experimental data of $\delta \alpha$ averaged over 20 temporal scans per point by a single-exponential decay function. The shaded regions represent the standard deviation of the exponential decay time obtained by fitting the measured dynamics of $\Delta \alpha$. The pre-excited electron density is $N=1.8\times10^{17}$ cm$^{-3}$ (black squares) and $N=1.5\times10^{18}$ cm$^{-3}$ (green circles) for the data in \textbf{a} and $N=4.4\times10^{15}$ cm$^{-3}$ (black squares) and $N=6.4\times10^{16}$ cm$^{-3}$ (green circles) for the data in \textbf{b}. }
		\label{Fig2}
	\end{figure}

	\begin{figure}[H]
		\center
		\includegraphics[width = 1\linewidth]{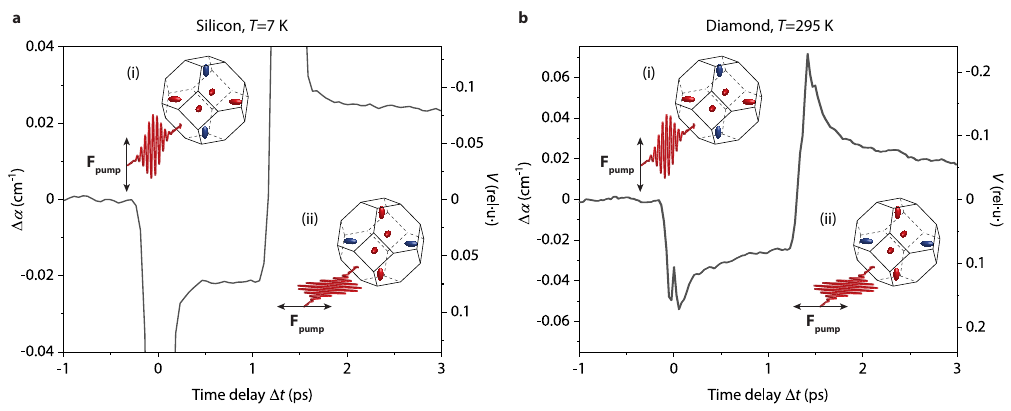}
		\captionsetup{labelformat=empty}
		\caption{\textbf{Fig. 3. Switching the valley polarization of electrons in silicon and diamond at THz frequency.} Valley polarized electron population is generated in silicon \textbf{a} at temperature of 7 K and in diamond \textbf{b} at room temperature using a pump pulse with linear polarization along [100] direction (insets (i) in \textbf{a} and \textbf{b}). After 1.4 ps, a second pump pulse with linear polarization along [010] direction (insets (ii) in \textbf{a} and \textbf{b}) switches the direction of valley polarization, which manifests itself as a change of sign of the measured $\Delta \alpha$ of the probe pulse, which is directly proportional to the degree of valley polarization $V$.}
		\label{Fig3}
	\end{figure}

	\section*{Methods}
	
	\subsection*{\label{sec:level2}Experimental setup}
	
	The infrared pump and probe pulses with photon energy of 0.62 eV, duration of 40 fs and repetition rate of 25 kHz are generated in a non-collinear optical parametric amplifier-difference frequency generation setup pumped by ytterbium femtosecond laser (Pharos SP, Light Conversion) \cite{Kozak2021}. Pulses with photon energy of 1.2 eV and duration of 170 fs (fundamental output from the laser) are used as pre-excitation pulses for the experiments in silicon. The spot size (1/$e^2$ radius) on the sample surface is $w_{\text{NIR}}=$ 73 $\mu$m. For experiments in diamond, the pre-excitation pulses with photon energy of 3.6 eV corresponding to the third harmonic frequency of the fundamental laser output are used. The beam is focused to the spot size of $w_{\text{UV}}=$ 57 $\mu$m. The peak excited carrier density is calculated from the measured absorbance of the pre-excitation pulses and the known area of laser foci. The time delays between the pre-excitation, the pump and the probe pulses are controlled using two independent optical delay lines.
	
	The infrared pump and probe pulses are focused to the sample using off-axis parabolic mirror with focal distance of $f=$ 100 mm. The spot size of both the pump and probe beams is $w_0=$ 26 $\mu$m. The direction of polarization of the pump pulse is controlled via broadband half wave plate. The sample is placed in a closed-cycle helium cryostat allowing to control its temperature in the range 7-295 K. The polarization anisotropy of free carrier absorption is measured by splitting the probe with linear polarization rotated by $\pm$45$^\circ$ with respect to the polarization of the pump pulse using a Glan-laser polarizer. The two polarization components of the probe parallel and perpendicular to the polarization of the pump pulse are detected by InGaAs photodiodes and the difference signal is measured by a lock-in amplifier. The amplifier is locked to the frequency of an optical chopper, which is placed into the pre-excitation beam. By this configuration we are only sensitive to the polarization anisotropy of free carrier absorption induced by the pump on the distribution of pre-excited carriers. Assuming that the relative change of the sample transmission induced by excited carriers is small, the absolute value of the polarization anisotropy of free carrier absorption $\Delta \alpha$ is calculated from the measured differential signal $S_{\text{A-B}}$ as $\Delta \alpha=S_{\text{A-B}}/(S_{\text{probe}}d)$, where $S_{\text{probe}}$ is the signal corresponding to the power of the transmitted probe pulse obtained by inserting the optical chopper to the probe beam and $d$ is the sample thickness.
	
	The two time separated pump pulses with orthogonal polarizations used in the switching experiments (results shown in Fig. 3) are generated by sending linearly polarized pump pulse with polarization rotated by 45$^\circ$ through an optically anisotropic crystal of beta-barium borate (BBO). This negative uniaxial anisotropic material introduces a group delay of 1.4 ps between the extraordinary and ordinary polarization components of the pump pulse polarized along [100] and [010] crystallographic directions of silicon and diamond crystals.
	
	\subsection*{\label{sec:level2}Numerical solution of Boltzmann transport equation using Monte-Carlo approach}
	
	To theoretically describe electron dynamics in momentum space during and after illumination of silicon and diamond crystals with the infrared pump pulse we numerically solve Boltzmann transport equation using Monte-Carlo approach following the treatment in \cite{Jacoboni1983}. This allows us to include both the anisotropic effective mass tensor of electrons in individual valleys and the energy and momentum dependent intra- and intervalley scattering mechanisms. We note that we only solve the equations for electrons because valence bands have their maxima in the $\Gamma$-point of the first Brillouin zone and the hole effective mass tensor is symmetric with respect to rotation by 90$^{\circ}$ in all valence bands (effective mass of holes has the same value along [100] and [010] directions). When we assume homogeneous distribution of electrons in real space, the time evolution of the momentum-dependent distribution function $f(\mathbf{k},t)$ can be written as:
	
	\begin{equation}
		\frac{\partial f(\mathbf{k},t)}{\partial t}-\frac{e\mathbf{F(t)}}{\hbar}\nabla_{\mathbf{k}}f(\mathbf{k},t)=\frac{\partial f(\mathbf{k},t)}{\partial t} \bigg|_{\text{coll}},
	\end{equation} where $\mathbf{F(t)}=\mathbf{F_0}\exp \left[-\frac{2\ln 2(t-t_0)^2}{\tau_\text{p}^2} \right]\cos \left ( \omega t\right )$ is the time dependent electric field of the pump pulse with the amplitude $\mathbf{F_0}$, angular frequency $\omega=2\pi f$ with $f$ being the central frequency of the pulse spectrum and FWHM duration of $\tau_\text{p}$, $e$ is the elementary charge and the right hand side represents the change of the distribution function due to collisions. The Monte-Carlo approach is based on solving the equation of motion for individual randomly chosen electrons from the initial distribution $f(\mathbf{k},0)$, where the collision term is expressed using a probability of electron scattering per unit volume and time, which has characteristic dependence on the electron energy/momentum for each type of scattering process. In our simulations we assume parabolic conduction band with six degenerate minima (valleys). Each valley possesses anisotropic effective mass tensor with the values of the longitudinal and transverse effective masses given in Table 1. We note that the parabolic approximation is sufficient to describe the generation and relaxation of valley polarization for our experimental parameters. The maximum kinetic energy which the electrons acquire during the oscillatory motion driven by the pump laser field is 250 meV in silicon ($F_0=0.7$ V/nm) and 0.6 eV in diamond ($F_0=1.3$ V/nm), which is much less than the depth of the valleys. The band nonparabolicity influences mainly the coherent nonlinear response of the electrons to the driving pump field but the intervalley scattering rates do not change significantly when taking the nonparabolicity into account. 
	
	To describe the electron collisions we assume several inelastic electron-phonon scattering mechanisms, namely the intravalley and intervalley electron scattering on acoustic and optical phonons. The probability per unit time of intravalley electron scattering on acoustic phonons from the initial state with momentum $\mathbf{k}$ to the final state with momentum $\mathbf{k}^\prime$ is:
	
	\begin{equation}
		P_{\text{a}}(\mathbf{k},\mathbf{k^\prime})=\frac{\pi q D^2_\text{a}}{V \rho u_l} \left[ \left( e^{\frac{\hbar q u_\text{l}}{k_\text{B}T}}-1 \right)^{-1}+\frac{1}{2} \pm \frac{1}{2}  \right] \delta \left( E_c(\mathbf{k^\prime})-E_c(\mathbf{k}) \pm \hbar q u_\text{l}\right),
	\end{equation} where $q$ is the length of the phonon wave vector, $D_\text{a}$ is deformation potential describing the inelastic interaction of electrons with acoustic phonons, $V$ is crystal volume, $\rho$ is material density, $u_\text{l}$ is propagation velocity of longitudinal acoustic phonons (speed of sound) in the material, $k_\text{B}$ is Boltzmann constant and $T$ is temperature. The intervalley electron scattering rate and the intravalley scattering rate at optical phonons is calculated as:
	
	\begin{equation}
		P_{\text{f(g)}}(\mathbf{k},\mathbf{k^\prime})=\frac{g \pi D^2_\text{f(g)}}{V \rho \Omega_{\text{f(g)}}} \left[ \left( e^{\frac{\hbar \Omega_{\text{f(g)}}}{k_\text{B}T}}-1 \right)^{-1}+\frac{1}{2} \pm \frac{1}{2}  \right] \delta \left( E_c(\mathbf{k^\prime})-E_c(\mathbf{k}) \pm \hbar \Omega_{\text{f(g)}} \right),
	\end{equation} where $D_\text{f(g)}$ is deformation potential describing the scattering of electrons at high energy optical or acoustic phonons leading to f-type (g-type) intervalley scattering, $g$ is the number of equivalent final valleys ($g=4$ for f-type intervalley optical phonon scattering to inequivalent valleys, $g=2$ for g-type optical phonon scattering to equivalent valleys) and $\Omega_{\text{f(g)}}$ is the frequency of the phonon involved in the intervalley scattering. In the simulations of electron dynamics in silicon we take into account f-type intervalley scattering at transverse and longitudinal acoustic (TA and LA) phonons and transverse optical (TO) phonons. In the simulations of electron dynamics in diamond we take into account g-scattering at longitudinal optical (LO) phonons and f-scattering at LA and TO phonons \cite{Jacoboni1983}. The parameters used in the simulations are summarized in Table 1. All the parameters except the intervalley deformation potentials are taken from \cite{Hammersberg2014}. For the deformation potentials, various values were reported in literature for both materials (see e.g. \cite{Isberg2013,Hammersberg2014,Jacoboni1983,Li2021} and references therein). The values of the acoustic deformation potential and the energies of phonons participating in f-type and g-type intervalley scattering in both materials are taken from \cite{Jacoboni1983}. The values of the deformation potential associated with the intervalley transitions from \cite{Jacoboni1983} lead to excessively long intervalley scattering times. For this reason we used deformation potentials as fitting parameters with the final values in a close agreement with the values reported in \cite{Isberg2013,Li2021}.
	
	\begin{table}[h]
		\centering
		\caption*{\textbf{Table 1: Parameters of the Monte-Carlo simulations}}
		\label{tab1}
		{\begin{tabular}{@{}llll@{}}
				\toprule
				Quantity & Silicon  & Diamond & Units\\
				\midrule
				$\rho$    & 2.329   & 3.530  & g/cm$^3$  \\
				$m_\text{l}$    & 0.92   & 1.56  & $m_0$  \\
				$m_\text{t}$    & 0.19   & 0.28  & $m_0$   \\
				$D_\text{a}$    &  9  &  8.7 & eV   \\
				$u_\text{l}$    &  8430  &  17520 & m/s   \\
				$D_\text{g}$    &    &  16 & $10^8$ eV/cm   \\
				$\hbar\Omega_{\text{g}}$    &    & 160  & meV   \\
				$D_\text{f1}$    & 0.6   &   & $10^8$ eV/cm   \\
				$\hbar\Omega_{\text{f1}}$    &  19  &   & meV   \\
				$D_\text{f2}$    & 4   & 16  & $10^8$ eV/cm   \\
				$\hbar\Omega_{\text{f2}}$    & 47   & 134  & meV   \\
				$D_\text{f3}$    & 4   & 16  & $10^8$ eV/cm   \\
				$\hbar\Omega_{\text{f3}}$    & 59   & 148  & meV   \\
				$F_0$    & 0.7   & 1.3  & V/nm   \\
				$\tau_\text{p}$    & 40   &  40  & fs   \\
				\bottomrule
		\end{tabular}}
	\end{table}
	
	\subsection*{\label{sec:level2}Polarization anisotropy of free carrier absorption described by Drude model}
	
	The free carrier absorption of infrared light due to free electrons and holes excited in the conduction and valence bands and relaxed to thermal equilibrium with the lattice can be described using Drude model, which represents a satisfactory approximation in the infrared spectral region. When assuming that the scattering rate is much lower than the frequency of the incident light and that the imaginary part of the dielectric function is small, the absorption induced by free electrons in a crystal for isotropic conduction band can be approximated as \cite{Yu1999}:
	
	\begin{equation}
		\label{Drude}
		\alpha_{\text{Drude}}=\frac{e^2N}{\varepsilon_0 m^* c n_0 \omega^2 \tau}=A \frac{N}{m^*},
	\end{equation} where $N$ is electron density, $\varepsilon_0$ is dielectric function of vacuum, $m^*$ is electron (hole) effective mass, $c$ is speed of light, $n_0$ is refractive index of the material, $\omega$ is light frequency, $\tau$ is electron scattering time and $A$ is the constant used in Eq. (1) in the main text. Eq. (\ref{Drude}) has to be modified to describe the total free carrier absorption including different populations of electrons in inequivalent valleys, anisotropic effective mass of electrons and the presence of holes. The total free carrier absorption coefficient in silicon and diamond for the infrared light linearly polarized along [100] and [010] directions can be written as:
	
	\begin{equation}
		\begin{array}{l}
			\label{Drude2}
			\alpha_{[100]}=\dfrac{e^2}{\varepsilon_0 c n_0 \omega^2 \tau} \left( \dfrac{N_{[100]}}{m_\text{l}} + \dfrac{N_{[010]}+N_{[001]}}{m_\text{t}} + \dfrac{N_{\text{h}}}{m_\text{h}}\right), \\
			\alpha_{[010]}=\dfrac{e^2}{\varepsilon_0 c n_0 \omega^2 \tau} \left( \dfrac{N_{[010]}}{m_\text{l}} + \dfrac{N_{[100]}+N_{[001]}}{m_\text{t}} + \dfrac{N_{\text{h}}}{m_\text{h}}\right),
			
		\end{array}
	\end{equation} where $N_{[100]}, N_{[010]}$ and $N_{[001]}$ are the real space electron densities in the valleys with principal axes along [100], [010] and [001] directions, \\$N_{\text{h}}=\left(N_{[100]}+N_{[010]}+N_{[001]} \right)$ is the density of holes and $m_\text{h}$ is the effective mass of holes (an average value over all bands when taking into account their occupation). Here we assume that the average carrier scattering time $\tau$ is isotropic and constant for electrons and holes. We note that the Drude model has been proven to describe the dielectric response of silicon in low frequency spectral region at nonequilibrium conditions \cite{Sato2014}. The polarization anisotropy of free carrier absorption (the result of Eq. (\ref{eq_abs})) is calculated as $\Delta\alpha=\alpha_{[100]}-\alpha_{[010]}$. The values of average electron scattering time $\tau$ used in the Drude model to calculate the absolute value of the polarization anisotropy of free carrier absorption are $\tau_\text{Si}=90$ fs in silicon and $\tau_\text{C}=30$ fs in diamond.
	
	The experimental value of $\Delta \alpha$ is determined from the measured absorption coefficients of the individual polarization components of the probe beam:
	
	\begin{equation}
		\label{absorption_coefficient}
		\alpha_{[100]([010])}=-\frac{\Delta T^{[100]([010])}}{T_0 d},
	\end{equation} where $\Delta T^{[100]([010])}$ corresponds to the change of transmission of the probe beam polarized along [100]([010]) direction induced by the excited charge carriers, $T_0$ is the probe transmission in an unexcited sample and $d$ is sample thickness. Here we assume that the free carrier absorption is weak and $\Delta T \ll T_0$.

	\subsection*{\label{sec:level2}Silicon and diamond crystals}
	
	Monocrystalline silicon sample used in this study was purchased from MicroChemicals GmbH. It is an intrinsic crystal without doping with low concentration of impurities and high resistivity of $\rho>$20000 $\Omega$.cm. The crystal was grown by vertical zone melting. The sample has a thickness of $d=$525 $\mu$m. The diamond crystal was manufactured by ElementSix using chemical vapor deposition. It has a low amount of impurities ($<1$ ppb of boron and $<5$ ppb of nitrogen as specified by the manufacturer). The crystal thickness is $d=$500 $\mu$m. Both samples have both sides polished and the surface normal oriented in [001] direction.
	
	\section*{Data availability}
	
	All the data that support the plots and the other findings of this study are publicly available at https://doi.org/10.5281/zenodo.15004810 \cite{Data}.
	
	\section*{Code availability}
	All the computational codes that were used to generate the data presented in this study are available from the corresponding author upon reasonable request.

	\section*{Methods-only references}
	
	38. Kozák, M., Peterka, P., Dostál, J., Trojánek, F. \& Malý, P. Generation of 
	
	few-cycle laser pulses at 2 $\mu$m with passively stabilized carrier-envelope phase 
	
	characterized by f-3f interferometry. \textit{Opt. Laser Technol.} \textbf{144}, 107394 
	
	(2021).
	
	\noindent
	39. Li, Z., Graziosi, P. \& Neophytou, N. Deformation potential extraction and 
	
	computationally efficient mobility calculations in silicon from first principles. 
	
	\textit{Phys. Rev. B} \textbf{104}, 195201 (2021).
	
	\noindent
	40. Sato, S. A., Shinohara, Y., Otobe, T. \& Yabana, K. Dielectric response of 
	
	laser-excited silicon at finite electron temperature. \textit{Phys. Rev. B} \textbf{90}, 174303
	
	(2014).

	\clearpage
	\begin{figure}
		\center
		\includegraphics[width = 1\linewidth]{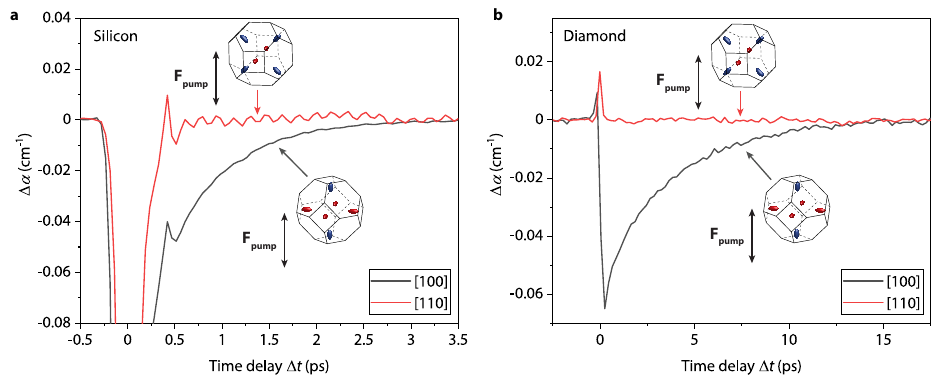}
		\captionsetup{labelformat=empty}
		\caption{\textbf{Extended Data Fig. 1. Polarization anisotropy of free carrier absorption in silicon and diamond.} Measured difference between the polarization components of free carrier absorption parallel and perpendicular to the pump field $\Delta \alpha$ in \textbf{a} silicon and \textbf{b} diamond for sample orientation with pump field along [100] direction (black curves) and [110] direction (red curves). The oscillations observed in the signal with pump polarization along [110] in silicon are due to impulsive excitation of coherent optical phonons with energy of 65 meV, which is smaller than the spectral width of the pump pulse, via stimulated Raman scattering. This process is not allowed with pump polarization along [100] due to symmetry reasons. Phonon oscillations are not present in diamond because the optical phonon energy of 160 meV is larger than the bandwidth of the pump pulse. Insets: Brillouin zone of silicon and diamond with the six conduction band valleys for the configuration with pump along [100] direction (lower insets) and [110] direction (upper insets). Blue ellipsoids correspond to the valleys with higher electron population after interacting with the pump pulse while red ellipsoids show valleys with lower electron population.}
		\label{extFig1}
	\end{figure}
	
	\clearpage
	\begin{figure}
		\center
		\includegraphics[width = 1\linewidth]{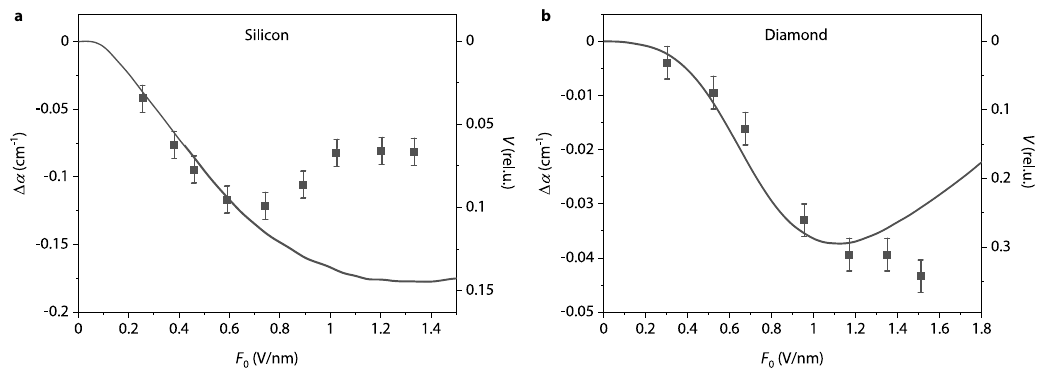}
		\captionsetup{labelformat=empty}
		\caption{\textbf{Extended Data Fig. 2. Dependence of polarization anisotropy of free carrier absorption in silicon and diamond on the electric field amplitude of the pump pulse.} Measured difference between the polarization components of free carrier absorption parallel and perpendicular to the pump field $\Delta \alpha$ and the associated degree of valley polarization $V$ in \textbf{a} silicon and \textbf{b} diamond (black squares) as a function of the electric field amplitude of the pump pulse $F_0$ compared with the results obtained by numerical Monte-Carlo simulations with the free carrier absorption described using Drude model with the electron density of $N_{\text{Si}}=7.5\times10^{17}$ cm$^{-3}$ in silicon and $N_{\text{C}}=2.6\times10^{16}$ cm$^{-3}$ in diamond. The values of the average electron scattering time in the two materials used in the Drude model are $\tau_\text{Si}=90$ fs and $\tau_\text{C}=30$ fs, respectively.}
		\label{extFig2}
	\end{figure}
	
	\clearpage
	\begin{figure}
		\center
		\includegraphics[width = 1\linewidth]{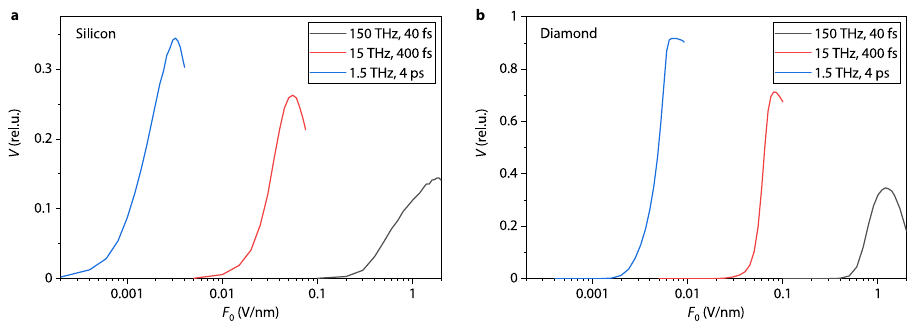}
		\captionsetup{labelformat=empty}
		\caption{\textbf{Extended Data Fig. 3. Numerical simulations of the degree of valley polarization induced in silicon and diamond at temperature of 77 K using pump pulses with different parameters.} The data show the calculated degree of valley polarization $V$ as a function of the peak electric field amplitude of the pump pulse in \textbf{a} silicon and \textbf{b} diamond for three different sets of pulse parameters, namely central photon energy of 0.62 eV (carrier frequency $f=$150 THz) and pulse duration 40 fs (black curves, pulse parameters applied in the experiments presented in this work), central photon energy of 62 meV ($f=$15 THz) and pulse duration of 400 fs (red curves) and central photon energy of 6.2 meV ($f=$1.5 THz) and pulse duration of 4 ps (blue curves).}
		\label{extFig3}
	\end{figure}

	\clearpage
	\section*{Supplementary Information:}
	\begin{figure}[!htbp]
		\centering
		\includegraphics[width = 0.9\linewidth]{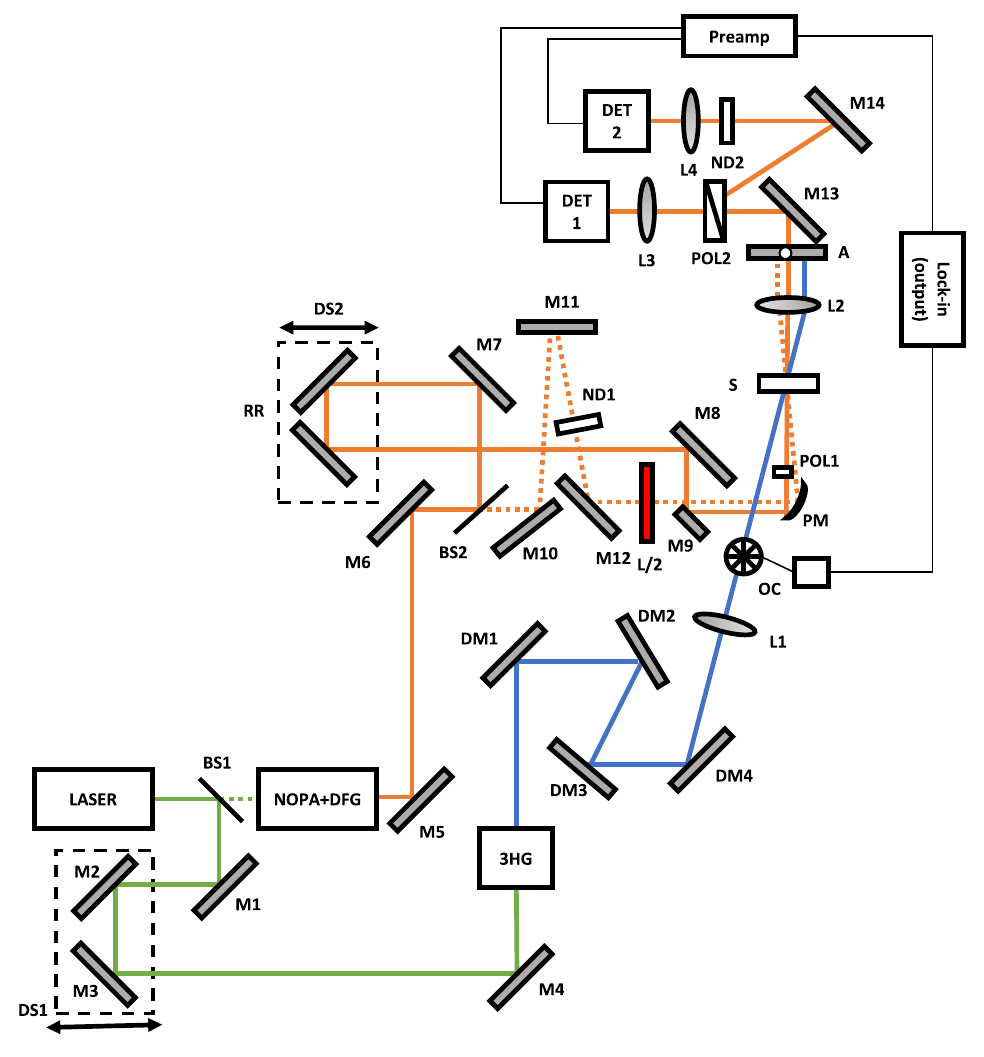}
		\caption{\textbf{Supplementary Fig. 1.} Detailed layout of the experimental setup. LASER - femtosecond laser system Pharos SP 6W (Light Conversion), NOPA+DFG - noncolinear optical parametric amplifier with subsequent difference frequency generation setup, where pump and probe pulses are generated, 3HG - third harmonic generation (used only during the experiments in diamond), BS1-2 - beam-splitter, M1-14 - silver mirror, DS1-2 - delay stage, DM1-4 - dielectric mirror, A - aperture, ND1-2 - neutral density filter, L1-4 - optical lens, OC - optical chopper, S - sample, RR - retroreflector, PM - parabolic mirror,
		POL1-2 - polarizer, L/2 - half-wave plate, DET1-2 - detector, Preamp - preamplifier, Lock-in - lock-in amplifier.}
		\label{FigureS1}
	\end{figure}
	
	\clearpage
	\textbf{Supplementary Video 1.} Monte Carlo simulation of electron dynamics during the interaction with the electric field of the infrared pump pulse in the first Brillouin zone of silicon at room temperature. The pump pulse with linear polarization along the [100] ($k_x$) direction accelerated the electron population in momentum space. The electrons that initially were in valleys with a large effective mass in the direction of the pump field are represented by blue dots. Electrons from valleys with a low effective mass are represented by red dots. During the interaction with the pump field, the lighter electrons mostly underwent intervalley scattering whereas the heavier electrons preferentially remained in the same valley.
	
	\textbf{Supplementary Video 2.} Monte Carlo simulation of electron dynamics during the interaction with the electric field of the infrared pump pulse in the first Brillouin zone of diamond at room temperature. The pump pulse with linear polarization along the [100] (kx) direction accelerated the electron population in momentum space. The electrons that initially were in valleys with a large effective mass in the direction of the pump field are represented by blue dots. Electrons from valleys with a low effective mass are represented by red dots. During the interaction with the pump field, the lighter electrons mostly underwent intervalley scattering whereas the heavier electrons preferentially remained in the same valley.

\end{document}